\documentclass[aps,prc,twocolumn,showpacs,superscriptaddress]{revtex4}
\usepackage{graphicx,color}% Include figure files

\usepackage{slashed}
\usepackage{amsmath}
\usepackage{amssymb,bm}
\usepackage{booktabs}

\begin{document}
% -------------------------------
\title{Higher-twist contributions to the transverse momentum broadening in semi-inclusive deep
inelastic scattering off large nuclei}

\author{Jia Jun Zhang}

\affiliation{Institute of Particle Physics, Central China Normal University, Wuhan 430079, China}
\affiliation{Key Laboratory of Quark and Lepton Physics (MOE), Central China Normal University, Wuhan 430079, China}
                  
\author{Jian-Hua Gao}
\affiliation{Shandong Provincial Key Laboratory of Optical Astronomy and Solar-Terrestrial Environment, Institute of Space Sciences, Shandong University, Weihai 264209, China}
 \affiliation{Key Laboratory of Quark and Lepton Physics (MOE), Central China Normal University, Wuhan 430079, China}

\author{Xin-Nian Wang}
%\email{xnwang@lbl.gov}
\affiliation{Institute of Particle Physics, Central China Normal University, Wuhan 430079, China}
\affiliation{Key Laboratory of Quark and Lepton Physics (MOE), Central China Normal University, Wuhan 430079, China}
\affiliation{Nuclear Science Division, MS 70R0319, Lawrence Berkeley National Laboratory, Berkeley, CA 94720}

\begin{abstract}
Multiple scattering leads to transverse momentum broadening of the struck quark in semi-inclusive deeply inelastic scatterings (SIDIS).
Nuclear broadening of the transverse momentum squared at the leading twist is determined by the twist-four collinear quark-gluon correlation function of the target nucleus that is in turn related to the jet transport parameter inside the nuclear medium. The twist-six contributions to the transverse momentum broadening are calculated as power corrections $\sim 1/Q^2$. Such power corrections are found to have no extra nuclear enhancement beyond the twist-four matrix elements and are determined by the nuclear modification of collinear parton distribution and correlation functions. They become important for an accurate extraction of the jet transport parameter inside large nuclei and its scale evolution at intermediate values of the hard scale $Q^2$.
\end{abstract}

\maketitle
% ========================================================
%%%%%%%%%%%%%%%%%%%%%%%%%%%%%%%%%%%%%%%%%%%%%%%%%%%%%%%%%%
\section{Introduction}

Multiple scattering of a propagating parton inside a medium leads to many interesting phenomena such as 
transverse momentum broadening and parton energy loss \cite{Gyulassy:1993hr,Baier:1996sk,Zakharov:1996fv,Gyulassy:2000fs,Wiedemann:2000za,Guo:2000nz}. These phenomena as observed in experiments in both high-energy heavy-ion collisions and semi-inclusive deeply inelastic scatterings (SIDIS) can in turn be used to extract properties of hot and cold nuclear matter. One of these medium properties as probed by propagating partons is the jet transport parameter $\hat q$. It is defined as the transverse momentum broadening squared per unit length \cite{Baier:1996sk} and measures the interaction strength between the energetic parton and the medium. A recent comprehensive phenomenological study \cite{Burke:2013yra} of experimental data on jet quenching in high-energy heavy-ion collisions at both the Relativistic Heavy-ion Collider (RHIC) and the Large Hadron Collider (LHC)  has extracted values of the jet transport parameter at the center of the produced dense matter. They are about two orders of magnitude higher than that in a cold nucleus as extracted from experimental data on semi-inclusive deeply inelastic scatterings (SIDIS) \cite{Deng:2009qb,Chang:2014fba}. Further improvements in the accuracy of the determination of the jet transport parameter depend on the study of parton energy loss and the transverse momentum broadening at the next-to-leading order \cite{Kang:2013raa} and evolution of the jet transport parameter \cite{CasalderreySolana:2007sw,Liou:2013qya,Blaizot:2014bha,Iancu:2014kga}.

Within the framework of generalized collinear factorization, multiple parton scattering and transverse momentum broadening can be expressed in terms of nuclear enhanced higher-twist contributions which depend on the twist-4 parton correlation functions \cite{Qiu:1990xy,Luo:1993ui,Guo:1998rd,Majumder:2007hx}. In the same framework, we can also calculate parton energy loss and the leading hadron suppression \cite{Guo:2000nz,Wang:2001ifa} in SIDIS. In this paper we make the first attempt to calculate the first power corrections to the transverse momentum broadening in SIDIS off a large nucleus. Similar to leading higher-twist corrections to hadron spectra, such power corrections to the transverse momentum broadening can become non-negligible for intermediate values of the hard scale $Q^2$. Inclusion of these higher-twist corrections should be important for a more accurate determination of the jet transport parameter from experimental data on transverse momentum broadening and its scale evolution.

Multiple soft interactions between the struck quark and the remnant nucleus in SIDIS can also be resummed in the eikonal limit into the gauge link in the transverse momentum dependent (TMD) parton distribution functions \cite{Ji:2004wu}. The transverse momentum broadening in a large nucleus arises naturally from the expectation value of the gauge link between the nuclear state. Under a maximal two-gluon correlation approximation, the transverse momentum broadening takes the form of a Gaussian distribution with the width given by the path-integrated jet transport parameter \cite{Liang:2008vz}. Within a systematic collinear expansion of the hard parts of partonic scattering, one can express the cross section of SIDIS in terms of TMD parton distribution and correlation functions with different power corrections \cite{Liang:2006wp}. Calculations of the SIDIS cross section have been carried out in the leading order and up to twist-4 power corrections \cite{Song:2010pf}. The results have been applied to the study of nuclear dependence of the azimuth asymmetry in both unpolarized \cite{Gao:2010mj} and polarized nuclear targets \cite{Song:2013sja,Song:2014sja}. We will employ these cross sections in this paper to study the transverse momentum broadening up to ${\cal O}(1/Q^2)$ corrections.

The remainder of this paper is organized as the following. In Section \ref{sec2}, we present the kinematics of SIDIS and review the
differential cross section of the SIDIS process in terms of the TMD parton distribution and correlation functions within the collinear expansion. In Section \ref{sec3},  we calculate the transverse momentum squared ($\vec k_\perp^2$) weighted differential SIDIS cross section. The corresponding $\vec k_\perp^2$-weighted parton distribution and correlation functions are shown to become higher-twist collinear (transverse momentum integrated) matrix elements which can be related to the jet transport parameter. We calculate the transverse momentum broadening up to ${\cal O}(1/Q^2)$ corrections in Section \ref{sec4}.  Conclusions and outlooks are given in Section \ref{sec5}.

%%%%%%%%%%%%%%%%%%%%%%%%%%%%%%%%%%%%%%%%%%%%%%%%%%%%%%%%%%%
\section{SIDIS cross section up to ${\cal O}(1/Q^2)$}
\label{sec2} 

We consider the differential cross section of the SIDIS process 
$e^-(\ell) + A(p) \rightarrow e^-(\ell')+q(k)+X(p_X)$ with an unpolarized
electron beam $e^-$ and an unpolarized target nucleus $A$ which carries momentum $p$ per nucleon in the infinite momentum frame,
\begin{eqnarray}
d\sigma&=&\frac{(4\pi\alpha_{em})^2}{2sQ^4}
\frac{d^3\ell'}{2E_{\ell'}(2\pi)^3}dW^{\mu\nu}L_{\mu\nu},
\end{eqnarray}
where $\alpha_{\rm em}$ is the fine structure constant in QED, $s$ is
the Mandelstam variable for the center-of-mass energy squared, $q=\ell-\ell'$ and $Q^2=-q^2$ is the
virtuality of the virtual photon that probes the nucleus. We neglect all the masses of particles.

\begin{figure}
\begin{center}
\includegraphics[width=8.0cm]{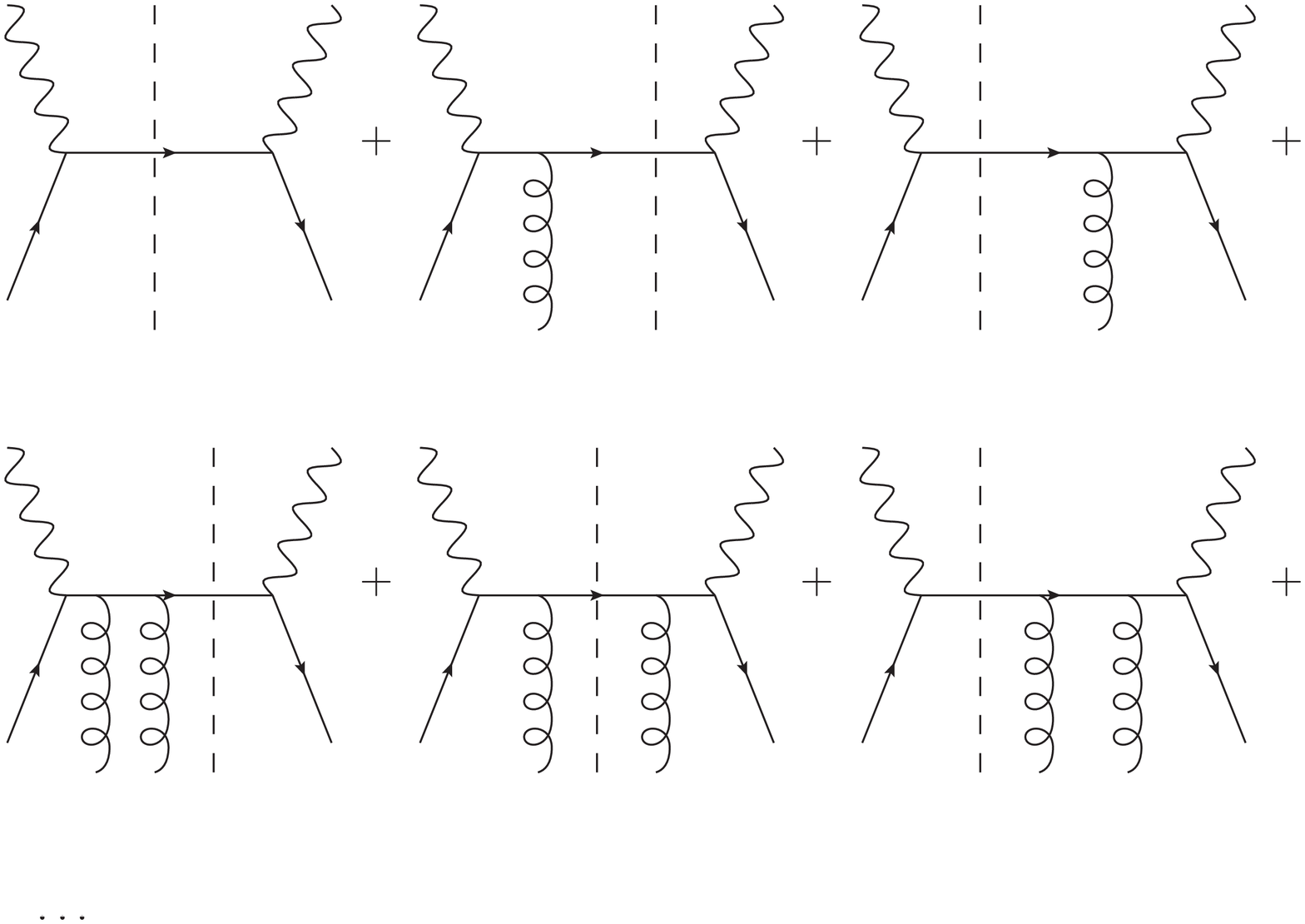}\\
\caption{Feynman diagrams for multiple scattering in SIDIS.}
\label{fig01}
\end{center}
\end{figure}

The leptonic tensor $L^{\mu\nu}$ and hadronic tensor $W^{\mu\nu}$  are defined as,
\begin{eqnarray}
L_{\mu\nu}&=&\mathrm{Tr}[\slashed{\ell}\gamma_{\mu}\slashed{\ell}'\gamma_{\nu}]
=4(\ell_{\mu}\ell'_{\nu}+\ell_{\nu}\ell'_{\mu}-\ell\cdot \ell' g_{\mu\nu}),\hspace{24pt}\\
W^{\mu\nu}&=&\frac{1}{2\pi}\sum_X\langle A|J_{\mu}(0)|X+q\rangle
\langle X+q|J_{\nu}(0)|A\rangle \nonumber\\
&&\hspace{0.5in}\times (2\pi)^4\delta^4(p+q-k-p_X),
\end{eqnarray}
respectively. In the infinite momentum frame, we parameterize the momenta in the following conventional light-cone 
expansion \cite{Liang:2006wp,Song:2010pf},
\begin{eqnarray}
p^{\mu}&=&p^+\bar{n}^{\mu},\\
\ell^{\mu}&=&\frac{1-y}{y}x_Bp^+\bar{n}^{\mu}+\frac{1}{y}q^-n^{\mu}
+\ell_T^{\mu},\\
\ell'^{\mu}&=&\frac{1}{y}x_Bp^+\bar{n}^{\mu}+\frac{1-y}{y}q^-n^{\mu}
+\ell_T^{\mu},\\
q^{\mu}&=&-x_Bp^+\bar{n}^{\mu}+q^-n^{\mu},
%
% k_1^{\mu}&=&x_1p^+\bar{n}^{\mu}+k_{1T}^{\mu},\\
%
% k_2^{\mu}&=&x_2p^+\bar{n}^{\mu}+k_{2T}^{\mu},\\
%
% d^{\mu\nu}&=&g^{\mu\nu}-\bar{n}^{\mu}n^{\nu}-\bar{n}^{\nu}n^{\mu}
% =g^{\mu\nu}-\bar{n}^{\{\mu}n^{\nu\}},\\
%
% \omega_{\rho}^{\ \rho'}&=&g_{\rho}^{\ \rho'}
 % -\bar{n}_{\rho}n^{\rho'},\\
%
% \epsilon_{\perp}^{\mu\nu}&=&\epsilon^{\mu\nu\rho\sigma}n_{\rho} \bar{n}_{\sigma},
\end{eqnarray}
where the unit vectors are taken as $\bar{n}^{\mu}=(1,0,0,0),
n^{\mu}=(0,1,0,0)$.
%n_{\perp1}^{\mu}=(0,0,1,0),n_{\perp2}^{\mu}=(0,0,0,1)$.

At the leading order (LO) of a perturbative expansion of the hard photon-parton scattering processes, one still has to
consider multiple interaction between the struck quark and the remanent nucleus through soft gluon
exchanges as shown in Fig.~\ref{fig01}.  The hadronic tensor can be written as a
sum of the contributions from all possible diagrams with multiple gluon exchanges. With collinear expansion of
the hard partonic parts, the hadronic tensor can be reorganized as products of collinear hard parts and gauge invariant TMD
quark (gluon) correlation functions~\cite{Ellis:1982wd,Ellis:1982cd,Qiu:1990xxa,Qiu:1990xy,Qiu:1991wg,Liang:2006wp}.
One can further reorganize the final differential cross section in terms of a power expansion in $1/Q$. 
The differential cross section of the unpolarized SIDIS process  $e^-(\ell) + A(p) \rightarrow e^-(\ell')+q(k)+X(p_X)$ 
up to the power of ${\cal O}(1/Q^2)$ takes the form \cite{Song:2010pf},
\begin{widetext}
\begin{eqnarray}
\frac{d\sigma^A}{dx_Bdyd^2\vec k_\perp}&=&\frac{2\pi\alpha^2_{\rm em}e_q^2}{Q^2y} \bigg\{[1+(1-y)^2]f_q^A(x_B,\vec k_\perp)
- 4(2-y)\sqrt{1-y}\frac{|\vec{k}_{\perp}|}{Q}x_B f_{q\perp}^A(x_B,\vec k_\perp)\cos\phi \notag\\
&& \hspace{-0.5in} -4(1-y)\frac{|\vec{k}_{\perp}|^2}{Q^2}x_B [\varphi^{(1)A}_{\perp2}(x_B,\vec k_\perp)
-\tilde{\varphi}^{(1)A}_{\perp2}(x_B,\vec k_\perp)]\cos2\phi
+16(1-y)\frac{x_B^2M^2}{Q^2} f^A_{q(-)}(x_B,\vec k_\perp)
\notag\\
&&\hspace{-0.5in} -2\left[1+(1-y)^2\right]\frac{|\vec{k}_{\perp}|^2}{Q^2}x_B
\left[\varphi^{(2)A}_{\perp}(x_B,\vec k_\perp)
-\tilde{\varphi}^{(2)A}_{\perp}(x_B,\vec k_\perp)\right]\bigg\},
\label{EQ01}
\end{eqnarray}
\end{widetext}
where the TMD parton distribution and correlation functions $f_q^A$, $f_{q\perp}^A$, $f_{q(-)}^A$,
$\varphi_{\perp2}^{(1)A}$, $\tilde{\varphi}_{\perp2}^{(1)A}$, $\varphi_{\perp}^{(2)A}$ and $\tilde{\varphi}_{\perp}^{(2)A}$ in terms of
matrix elements can be found in Eqs.~\eqref{EQA01}--\eqref{EQA13} in the appendix. The above expansion in terms of
TMD parton distribution and correlation functions is a generalized expansion similar to the collinear twist expansion. 
The first term is the leading twist contribution proportional to the TMD quark distribution function $f_q^A(x_B, \vec k_\perp)$. This
gives rise to the normal collinear quark distribution function after integration over the transverse momentum. The second
term is the twist-3 contribution with an azimuthal asymmetry $\cos\phi$. The rest are twist-four contributions.
All high-twist contributions are suppressed by $\mathcal{O}([|\vec k_\perp|/Q]^n, [M/Q]^n)$.

%%%%%%%%%%%%%%%%%%%%%%%%%%%%%%%%%%%%%%%%%%%%%%%%%%%%%%%%%%%%%%%%%%
\section{\label{sec3} Nuclear dependence of higher-twist matrix elements}

To study the transverse momentum broadening, we need to calculate the $\vec k_{\perp}^2$-weight-integrated differential cross section. Multiplying with $\vec k_\perp^2$ and integrating over the transverse momentum, one obtains from Eq.~(\ref{EQ01}),
\begin{eqnarray}
 \frac{d \langle \vec k_\perp^2 \sigma^A\rangle}{dx_Bdy}&=&\frac{2\pi\alpha^2_{\rm em}e^2_q}{Q^2y}
\bigg\{\left[1+(1-y)^2\right] \langle \vec k_\perp^2f^A_q\rangle (x_B) \notag\\
& + &\frac{16(1-y)x_B^2M^2}{Q^2} \langle \vec k_\perp^2 f^A_{q(-)} \rangle (x_B) \notag\\
 &-& \frac{2[1+(1-y)^2]x_B}{Q^2} \notag\\
 &&\hspace{-0.4in} \times \left[\langle \vec k_\perp^4\varphi_{\perp}^{(2)A}\rangle (x_B)
-\langle \vec k_\perp^4\tilde{\varphi}_{\perp}^{(2)A}\rangle (x_B)\right]\bigg\},
\label{EQ02}
\end{eqnarray}
where $\langle \vec k_\perp^2f^A_q\rangle (x_B)$, $ \langle \vec k_\perp^2 f^A_{q(-)} \rangle (x_B)$, $\langle \vec k_\perp^2 \varphi_{\perp}^{(2)A}\rangle (x_B)$ and $\langle \vec k_\perp^2 \tilde{\varphi}_{\perp}^{(2)A}\rangle (x_B)$ are $\vec k_\perp^2$-weight-integrated parton distribution/correlation functions of $f_q^A(x_B, \vec k_\perp)$, $f_{q(-)}^A(x_B,\vec k_\perp)$, $\varphi_{\perp}^{(2)A}(x_B,\vec k_\perp)$ and $\tilde{\varphi}_{\perp}^{(2)A}(x_B,\vec k_\perp)$, respectively [see Eqs.~\eqref{EQA14}--\eqref{EQA20} in the appendix]. These $\vec k_\perp^2$-weight-integrated parton distribution and correlation functions contain $\vec k_\perp^2$-weighted matrix elements. Through partial integration in the transverse momentum $\vec k_\perp$, one can convert each $\vec k_\perp$ into a partial derivative with respect to the transverse coordinate. The field operators in $\vec k_\perp^2$-weight-integrated parton distribution and correlation functions are always two twist higher than those in the corresponding distribution and correlation functions without $\vec k_\perp^2$-weight. Therefore the above $\vec k_\perp^2$-weighted cross section should contain parton correlation functions up to twist-six.

Take the first term in Eq.~\eqref{EQ02} as an example, the $\vec k_\perp^2$-weight-integrated parton distribution function,
\begin{eqnarray}
\langle \vec k_\perp^2f^A_q\rangle (x)& \equiv & \int d^2k_\perp f_q^A(x,\vec k_\perp) \vec k_\perp^2 \notag\\
&& \hspace{-0.7in}= -\int\frac{dy^-d^2y_{\perp}d^2\vec k_\perp}{(2\pi)^3}
e^{ixp^+y^--i\vec{k}_{\perp}\cdot\vec{y}_{\perp}} \notag\\
& & \hspace{-0.7in} \times \langle A|\bar{\psi}(0)\frac{\gamma^+}{2}(-i\partial_{\perp\rho})
(-i\partial_{\perp}^{\rho})\mathcal{L}(0;y)\psi(y)
|A\rangle\notag\\
&& \hspace{-0.7in} \simeq -\int \frac{dy^-}{2\pi}\int_{y^-}^{\infty}d\xi^-\int_{\xi^-}^{\infty} d\eta^- e^{ixp^+y^-} \notag\\
&& \hspace{-0.7in}\times 2g^2\langle A|\bar{\psi}(0) \frac{\gamma^+}{2}F_{\perp\rho}^{\ +}(\eta^-)
F_{\perp}^{\rho+}(\xi^-)\psi(y^-)|A\rangle,
%\notag\\
%&\simeq&A\int dy^- e^{ixp^+y^-}\langle N|\bar{\psi}(0)
%\frac{\gamma^+}{4\pi}\psi(y^-)|N\rangle\notag\\
%&&\times(-2g^2)\int_{y^-}^{\infty}d\xi^-\int_{\xi^-}^{\infty}
%d\eta^-\langle N|F_{\perp\rho}^{\ +}(\eta^-)F_{\perp}^{\rho+}(\xi^-)
%|N\rangle
 \label{EQ03}
\end{eqnarray}
is in fact a twist-four two-parton correlation function. In the above equation, we have used the identity \cite{Liang:2008vz},
\begin{widetext}
\begin{equation}
-i\partial_{\perp}^{\rho}\mathcal{L}(\infty;y^-,\vec{y}_\perp)=\mathcal{L}(\infty;y^-, \vec{y}_\perp)
\left[ D_{\perp}^{\rho}(y)+g\int_{y^-}^{\infty}d\xi^-
\mathcal{L}(y^-;\xi^-,\vec{y}_{\perp})
F_{\perp}^{\rho+}(\xi^-,\vec{y}_{\perp})
\mathcal{L}(\xi^-;y^-,\vec{y}_{\perp})\right],
\label{EQ04}
\end{equation}
\end{widetext}
for a derivative acting on a gauge link,
\begin{eqnarray}
\mathcal{L}(z^-;y^-,\vec{y}_\perp)={\cal P}\exp\left[-ig \int^{z^-}_{y^-} d\xi^- A^+(\xi^-, \vec{y}_\perp)\right].
\end{eqnarray}
The covariant derivative is defined as $D_\perp^\rho(y)\equiv -i\partial_\perp^\rho +gA_\perp^\rho(y)$. Note that the gauge link in Eq.~(\ref{EQ03}) between two space-time points with a transverse displacement is defined as,
\begin{equation}
\mathcal{L}(0;y)=\mathcal{L}^\dagger(\infty;0) \mathcal{L}(\infty;y^-,\vec{y}_\perp),
\end{equation}
in a covariant gauge. According to the above identity, the double derivatives on the gauge link in Eq.~(\ref{EQ04}) produce three different contributions.  We only keep the term with two gluon field strength operators which is enhanced by the nuclear size with a factor of $A^{4/3}$ due to the double path integrations across the nuclear size. Other terms that have one or two covariant derivatives on the quark field are ignored because they do not have such a nuclear enhancement.  We also omit the gauge link in the final expression of the parton correlation function for brevity. Note that one can change the limits of the integration over $\eta^-$ to the same as $\xi^-$ in Eq.~(\ref{EQ03}) with an overall factor of 1/2. 

Assuming a large and loosely bound nucleus, one can factorize the twist-four quark-gluon correlation function as a product of the quark distribution function for $A$ nucleons and soft gluon distribution density integrated over the size of the nucleus \cite{Osborne:2002st},
\begin{eqnarray}
& &\hspace{-24pt} \langle A|\bar{\psi}(0) \frac{\gamma^+}{2}F_{\perp\rho}^{\ +}(\eta^-) F_{\perp}^{\rho+}(\xi^-)\psi(y^-)|A\rangle \notag\\
&\approx & A  \langle N|\bar{\psi}(0) \frac{\gamma^+}{2} \psi(y^-) |N\rangle \frac{1}{2N_c} \frac{\rho_N(\xi_N^-)}{2p^-} \notag\\
&\times& \langle N| F_{\perp\rho}^{\ +}(\xi_N^- -\frac{\lambda^-}{2}) \
F_{\perp}^{\rho+}(\xi_N^-+\frac{\lambda^-}{2}) |N\rangle,
\end{eqnarray}
where $\xi_N^-=(\eta^-+\xi^-)/2$ and $\lambda^- =\xi^- - \eta^-$. We have taken the medium ensemble average,
\begin{equation}
\int \frac{d^3p}{(2\pi)^32p^+}f_A(p,\xi_N)\langle N| \cdots |N\rangle =\frac{\rho_N^A(\xi_N^-) }{2p^+}\langle N| \cdots |N\rangle \notag,
\end{equation}
for the gluon field strengths over all possible nucleons inside the nucleus and $\rho_N^A(\xi_N^-)$ is the spatial nucleon density normalized to the atomic number $A$.  Given the definition of the collinear quark and gluon distribution functions,
\begin{eqnarray}
f_q^N(x)&=& \int \frac{dy^-}{2\pi} e^{ixp^+y^-} \langle N|\bar{\psi}(0) \frac{\gamma^+}{2} \psi(y^-) |N\rangle,  \\
xg_N(x)&=&- \int \frac{d\lambda^-}{2\pi p^+} e^{ixp^+\lambda^-} 
\langle N| F_{\perp\rho}^{\ +}(0) F_{\perp}^{\rho+}(\lambda^-) |N\rangle, \notag \\
\end{eqnarray}
the $\vec k_\perp^2$-weight-integrated quark distribution function becomes
\begin{equation}
\langle \vec k_\perp^2f^A_q\rangle (x) =A f_q^N(x) \int d\xi_N^- \hat q_F(\xi_N^-) +{\cal O}(A),
\end{equation}
where 
\begin{equation}
\hat q_F(\xi_N^-) = \frac{2\pi^2\alpha_{\rm s}}{N_c} \rho_N^A(\xi_N^-)[x_g g_N(x_g)]_{x_g\approx0},
\end{equation}
is the jet transport parameter for a quark. It represents the transverse momentum broadening squared per unit length. The averaged total
transverse momentum broadening squared inside a nucleus is then,
\begin{equation}
\Delta_{2F}= \int d\xi_N^- \hat q_F(\xi_N^-),
\end{equation}
which should be proportional to the nuclear size $R_A\approx 1.12 A^{1/3}$ fm.

One can obtain similar relation for the $\vec k_\perp^2$-weight-integrated parton distribution function in the  second term  of Eq.~(\ref{EQ02}),
\begin{equation}
 \langle \vec k_\perp^2 f^A_{q(-)} \rangle (x)=A f_{q(-)}^N(x) \Delta_{2F}  +{\cal O}(A),
\end{equation}
where 
\begin{equation}
f^A_{q(-)}(x,\vec k_\perp)=\frac{p^+}{M^2}\int\frac{dy^-}
{2\pi}e^{ixp^+y^-} \langle A|\bar{\psi}(0)\frac{\slashed{p}}{2}
\psi(y^-)|A\rangle,
\end{equation}
is a twist-4 collinear quark distribution function.

Since the integration over the length of the nucleus should be proportional to the nuclear size $R_A\sim A^{1/3}$, the nuclear enhanced parts of both $\langle \vec k_\perp^2f^A_q\rangle (x)$ and $ \langle \vec k_\perp^2 f^A_{q(-)} \rangle (x)$ are proportional to $A^{4/3}$. One can neglect other contributions that are not enhanced by the nuclear size, including those that describe quark-gluon correlation functions inside a single nucleon.

The third term of Eq.~(\ref{EQ02}) contains contributions from double scattering processes in Fig.~\ref{fig01} with two gluon exchanges. The associated parton correlation functions have two covariant derivatives and have contributions from both left and right-cut diagrams,
\begin{equation}
\varphi_{\perp}^{(2)A}(x,\vec k_\perp)=\frac{1}{2}\left[\varphi_{\perp}^{(2,L)A}(x,\vec k_\perp)+\varphi_{\perp}^{(2,R)A}(x,\vec k_\perp) \right].
 \end{equation}
 The corresponding $\vec k_\perp^4$-weight-integrated parton correlation functions can be similarly simplified assuming factorization in a large and loosely bound nucleus,
 \begin{widetext}
\begin{eqnarray}
\langle \vec k_\perp^4\varphi_{\perp}^{(2,L)A}\rangle(x)&=&
%\int\frac{dx_2}{x_2-x-i\epsilon}
%\frac{dy^-d^2y_{\perp}d^2k_{\perp}}{(2\pi)^3}\frac{p^+dz^-}{2\pi}
%e^{ix_2p^+z^-+ixp^+(y^--z^-)-i\vec{k}_{\perp}\cdot\vec{y}_{\perp}}
%\left(d^{\rho\sigma}\vec{k}^2_{\perp}\right)\notag\\
%&&\times\langle A|\bar{\psi}(0)\frac{\gamma^+}{2}
%\mathcal{L}(0;z^-,0_{\perp})D_{\rho}(z^-,0_{\perp})
%D_{\sigma}(z^-,0_{\perp})\mathcal{L}(z^-,0_{\perp};y)\psi(y)|A\rangle,
%\notag\\
%&=&
\int\frac{dy^-d^2y_{\perp}d^2\vec k_\perp p^+dz^-}{(2\pi)^3}
i\theta(z^-)e^{ixp^+y^--i\vec{k}_{\perp}\cdot\vec{y}_{\perp}} \vec k_\perp^2 \notag\\
&&\times\langle A|\bar{\psi}(0)\frac{\gamma^+}{2}
\mathcal{L}(0;z^-,0_{\perp})D_{\perp\sigma}(z^-,0_{\perp})
D_\perp^\sigma(z^-,0_{\perp})\mathcal{L}(z^-,0_{\perp};y)
\psi(y)|A\rangle\notag\\
&\simeq&-\int\frac{dy^-p^+dz^-}{2\pi}
i\theta(z^-)e^{ixp^+y^-}\int_{y^-}^{\infty}d\xi^-
\int_{\xi^-}^{\infty}d\eta^- 2g^2\notag\\
&&\times\langle A|\bar{\psi}(0)\frac{\gamma^+}{2}
D_{\perp\sigma}(z^-,0_{\perp})D_{\perp}^{\sigma}(z^-,0_{\perp})
F_{\perp\rho}^{\ +}(\eta^-,0_{\perp})F_{\perp}^{\rho+}(\xi^-,0_{\perp})
\psi(y^-)|A\rangle,
\label{EQ09}\\
%\end{widetext}
%
% \begin{widetext}
\langle \vec k_\perp^4\varphi_{\perp}^{(2,R)A}\rangle(x)
%&=&\int\frac{dx_1}{x_1-x+i\epsilon}
%\frac{dy^-d^2y_{\perp}d^2k_{\perp}}{(2\pi)^3}\frac{p^+dz^-}{2\pi}
%e^{ixp^+z^-+ix_1p^+(y^--z^-)-i\vec{k}_{\perp}\cdot\vec{y}_{\perp}}
%\left(d^{\rho\sigma}\vec{k}_{\perp}^2\right)\notag\\
%&&\times\langle A|\bar{\psi}(0)\frac{\gamma^+}{2}
%\mathcal{L}(0;z^-,y_{\perp})D_{\rho}(z^-,y_{\perp})
%D_{\sigma}(z^-,y_{\perp})\mathcal{L}(z^-,y_{\perp};y)
%\psi(y)|A\rangle\notag\\
&=&-\int\frac{dy^-d^2y_{\perp}d^2\vec k_\perp p^+dz^-}{(2\pi)^3}
i\theta(z^--y^-)e^{ixp^+y^--i\vec{k}_{\perp}\cdot\vec{y}_{\perp}} \vec k_\perp^2\notag\\
&&\times\langle A|\bar{\psi}(0)\frac{\gamma^+}{2}
\mathcal{L}(0;z^-,y_{\perp})D_{\rho}(z^-,y_{\perp})
D_{\sigma}(z^-,y_{\perp})\mathcal{L}(z^-,y_{\perp};y)
\psi(y)|A\rangle\notag\\
&\simeq&\int\frac{dy^-p^+dz^-}{2\pi}
i\theta(z^--y^-)e^{ixp^+y^-}\int_0^{\infty}d\xi^-
\int_{\xi^-}^{\infty}d\eta^- 2g^2\notag\\
&&\times\langle A|\bar{\psi}(0)\frac{\gamma^+}{2}
F_{\perp}^{\rho+}(\xi^-,0_{\perp})F_{\perp\rho}^{\ +}(\eta^-,0_{\perp})
D_{\perp\sigma}(z^-,0_{\perp})D_{\perp}^{\sigma}(z^-,0_{\perp}),
\psi(y^-)|A\rangle,
\label{EQ10}
\end{eqnarray}
where we again omit the gauge links in the final expression for brevity. 
% We also use a little trick in Eq.~\eqref{EQ10} to simplify our calculation.
%We shift the coordinate frame along the transverse direction so that
%$y_{\perp}\rightarrow0_{\perp}$ and $0_{\perp}\rightarrow-y_{\perp}$,
%in this way we can make the partial derivatives operate only on the
%first gauge link in Eq.~\eqref{EQ10}.
Using Eq.~\eqref{EQ04} we can also get, 
\begin{eqnarray}
&& D_{\perp}^{\sigma}(z^-,\vec{y}_{\perp})
D_{\perp}^{\rho}(z^-,\vec{y}_{\perp})
\mathcal{L}(z^-;y^-,\vec{y}_{\perp})\notag\\
&& \hspace{24pt}\simeq g^2\int_{y^-}^{z^-}d\xi^-\int_{\xi^-}^{z^-}d\eta^-
\left[F_{\perp}^{\sigma+}(\eta^-,\vec{y}_{\perp})
F_{\perp}^{\rho+}(\xi^-,\vec{y}_{\perp})+
F_{\perp}^{\rho+}(\eta^-,\vec{y}_{\perp})
F_{\perp}^{\sigma+}(\xi^-,\vec{y}_{\perp})\right], \\
&&\mathcal{L}(0;z^-,\vec{y}_{\perp})
\overleftarrow{D}_{\perp}^{\sigma}(z^-)
\overleftarrow{D}_{\perp}^{\rho}(z^-)\notag\\
&& \hspace{24pt}\simeq g^2\int_0^{z^-}d\xi^-\int_{\xi^-}^{z^-}d\eta^-
\left[F_{\perp}^{\sigma+}(\xi^-,\vec{y}_{\perp})
F_{\perp}^{\rho+}(\eta^-,\vec{y}_{\perp})+
F_{\perp}^{\rho+}(\xi^-,\vec{y}_{\perp})
F_{\perp}^{\sigma+}(\eta^-,\vec{y}_{\perp})\right],
\label{EQ12}
\end{eqnarray}
%Note that in Eq.~\eqref{EQ12} the covariant derivatives operate on the left side rather than the right side. 
which help to reduce the $\vec k_\perp^4$-weight-integrated parton correlation functions to the following forms,
\begin{eqnarray}
\langle \vec k_\perp^4\varphi_{\perp}^{(2,L)A}\rangle(x)
&\simeq&\int\frac{dy^-p^+dz^-}{2\pi}
i\theta(z^-)e^{ixp^+y^-}\int_{y^-}^{\infty}d\xi^-
\int_{\xi^-}^{\infty}d\eta^-\int_0^{z^-}d\xi'^-
\int_{\xi'^-}^{z^-}d\eta'^-\notag\\
&&\times(-4g^4)\langle A|\bar{\psi}(0)\frac{\gamma^+}{2}
F_{\perp\sigma}^{\ +}(\xi'^-)F_{\perp}^{\sigma+}(\eta'^-)
F_{\perp\rho}^{\ +}(\eta^-)F_{\perp}^{\rho+}(\xi^-)
\psi(y^-)|A\rangle,\label{EQ13}\\
\langle \vec k_\perp^4\varphi_{\perp}^{(2,R)A}\rangle(x)
&\simeq&\int\frac{dy^-p^+dz^-}{2\pi}
i\theta(z^--y^-)e^{ixp^+y^-}\int_0^{\infty}d\xi^-
\int_{\xi^-}^{\infty}d\eta^-\int_{y^-}^{z^-}d\xi'^-
\int_{\xi'^-}^{z^-}d\eta'^-\notag\\
&&\times 4g^4 \langle A|\bar{\psi}(0)\frac{\gamma^+}{2}
F_{\perp}^{\sigma+}(\xi^-)F_{\perp\sigma}^{\ +}(\eta^-)
F_{\perp}^{\rho+}(\eta'^-)F_{\perp\rho}^{\ +}(\xi'^-)
\psi(y^-)|A\rangle.\label{EQ14}
\end{eqnarray}
\end{widetext}
These are twist-six matrix elements for parton correlation functions inside a nucleus. Under the same approximation of
a large and loosely bound nucleus, one can similarly factorize these twist-six matrix elements into products of twist-four or 
twist-two matrix elements in a general form,
\begin{eqnarray}
\langle \vec k_\perp^4\varphi_{\perp}^{(2,L/R)A}\rangle (x)&\simeq&A^{5/3} 
\langle\bar{\psi}\psi\rangle_N \otimes
\langle F F \rangle_N \otimes \langle F F \rangle_N\notag\\
&+&A^{4/3} \langle\bar{\psi}FF\psi\rangle _N \otimes
\langle F F \rangle_N\notag\\
&& \hspace{-0.5in} +A^{4/3} \langle FFFF \rangle _N
\otimes \langle\bar{\psi}\psi \rangle_N + {\cal O}(A),
\label{EQ15}
\end{eqnarray}
where $\langle \dots\rangle_N$ represents the expectation value over a nucleon state.
Since the hard photon-quark scattering can happen in any nucleon in the nucleus, summation of these interactions over all nucleons
inside the nucleus contributes to a factor $A$. Integration over the position of another nucleon in the secondary interaction along
the propagation path of the struck quark will give another factor of $R_A\sim A^{1/3}$. These matrix elements are enhanced by a factor of $A^{1/3}$ for each additional nucleons involved in the interaction beside the hard photon-quark scattering. Other contributions without the nuclear enhancement can be neglected here.

%We find that the leading contribution ($\mathcal{O}(A^{5/3})$)
%in Eq.~\eqref{EQ15} will finally cancel each other when we sum
%$\Psi_{\perp}^{(2,L)A}(x)$ and $\Psi_{\perp}^{(2,R)A}(x)$,
%the finite contributions come from the second and the third
%terms in Eq.~\eqref{EQ15}, that is $\mathcal{O}(A^{4/3})$.
%We will show the result below explicitly.

Let us exam separately contributions from the left and right-cut diagrams in the first term in Eq.~\eqref{EQ15} with the most nuclear enhancement, 
\begin{widetext}
\begin{eqnarray}
\langle \vec k_\perp^4\varphi_{\perp}^{(2,L)A}\rangle_1(x)&\simeq&
-A\int\frac{dy^-}{2\pi}e^{ixp^+y^-}\langle N|\bar{\psi}(0)
\frac{\gamma^+}{2}\psi(y^-)|N\rangle \int_0^{\infty}d\xi^-\int_{\xi^-}^{\infty}d\eta^-
2g^2\langle N|F_{\perp\rho}^{\ +}(\eta^-)F_{\perp}^{\rho+}(\xi^-)
|N\rangle\notag\\
&&\times\int p^+dz^-i\theta(z^-)\int_0^{z^-}d\xi'^-
\int_{\xi'^-}^{z^-}d\eta'^-2g^2\langle N|
F_{\perp\sigma}^{\ +}(\xi'^-)F_{\perp}^{\sigma+}(\eta'^-)|N\rangle \frac{\rho_N^A(\xi_N^-)}{2p^-}\frac{\rho_N^A({\xi'}_N^-)}{2p^-}
 \label{EQ16}
\end{eqnarray}
\end{widetext}
\begin{widetext}
\begin{eqnarray}
\langle \vec k_\perp^4\varphi_{\perp}^{(2,R)A}\rangle_1(x) &\simeq&A\int\frac{dy^-}{2\pi}
e^{ixp^+y^-}\langle N|\bar{\psi}(0)\frac{\gamma^+}{2}
\psi(y^-)|N\rangle \int_0^{\infty}d\xi^-\int_{\xi^-}^{\infty}d\eta^-
2g^2\langle N|F_{\perp}^{\sigma+}(\xi^-)
F_{\perp\sigma}^{\ +}(\eta^-)|N\rangle\notag\\
&&\times\int p^+dz^-i\theta(z^--y^-)\int_{y^-}^{z^-}
d\xi'^-\int_{\xi'^-}^{z^-}d\eta'^-2g^2\langle N|
F_{\perp}^{\rho+}(\eta'^-)F_{\perp\rho}^{\ +}(\xi'^-)
|N\rangle \frac{\rho_N^A(\xi_N^-)}{2p^-}\frac{\rho_N^A({\xi'}_N^-)}{2p^-}
\notag\\
%&=&A\int\frac{dy^-}{2\pi}e^{ixp^+y^-}\langle N|\bar{\psi}(0)
%\frac{\gamma^+}{2}\psi(y^-)|N\rangle\notag\\
%&&\times\int_0^{\infty}d\xi^-\int_{\xi^-}^{\infty}d\eta^-
%2g^2\langle N|F_{\perp}^{\sigma+}(\xi^-)F_{\perp\sigma}^{\ +}(\eta^-)
%|N\rangle\notag\\
%&&\times\int p^+dz^-i\theta(z^--y^-)\int_{y^-}^{z^-}d\xi'^-
%\int_{y^-}^{\xi'^-}d\eta'^-2g^2\langle N|F_{\perp}^{\rho+}(\xi'^-)
%F_{\perp\rho}^{\ +}(\eta'^-)|N\rangle\notag\\
%&=&A\int\frac{dy^-}{2\pi}e^{ixp^+y^-}\langle N|\bar{\psi}(0)
%\frac{\gamma^+}{2}\psi(y^-)|N\rangle\notag\\
%&&\times\int_0^{\infty}d\xi^-\int_{\xi^-}^{\infty}d\eta^-
%2g^2\langle N|F_{\perp}^{\sigma+}(\xi^-)F_{\perp\sigma}^{\ +}(\eta^-)
%|N\rangle\notag\\
%&&\times\int p^+dz^-i\theta(z^--y^-)\int_0^{z^--y^-}d\xi'^-
%\int_0^{\xi'^-}d\eta'^-2g^2\langle N|F_{\perp}^{\rho+}(\xi'^-)
%F_{\perp\rho}^{\ +}(\eta'^-)|N\rangle\notag\\
&=&A\int\frac{dy^-}{2\pi}e^{ixp^+y^-}\langle N|\bar{\psi}(0)
\frac{\gamma^+}{2}\psi(y^-)|N\rangle \int_0^{\infty}d\xi^-\int_{\xi^-}^{\infty}d\eta^-
2g^2\langle N|F_{\perp}^{\sigma+}(\xi^-)F_{\perp\sigma}^{\ +}(\eta^-)
|N\rangle\notag\\
&&\times\int p^+dz^-i\theta(z^-)\int_0^{z^-}d\xi'^-
\int_0^{\xi'^-}d\eta'^-2g^2\langle N|F_{\perp}^{\rho+}(\xi'^-)
F_{\perp\rho}^{\ +}(\eta'^-)|N\rangle \frac{\rho_N^A(\xi_N^-)}{2p^-}\frac{\rho_N^A({\xi'}_N^-)}{2p^-},
 \label{EQ17}
\end{eqnarray}
\end{widetext}
where $\xi_N^-=(\eta^- + \xi^-)/2$ and ${\xi'}_N^-=(\eta'^- +\xi'^-)/2$ are the light-cone positions of the nucleons. In Eq.~\eqref{EQ16} we approximate the integral boundary for $\xi^-\in (y^-,\infty)$ by  $\xi^-\in (0,\infty)$. 
Since $y^-$ is confined to the size of a nucleon for large and moderate values of momentum fraction $x$, the difference should be small.
In Eq.~\eqref{EQ17} we first interchange the integration variables $\eta'^- \leftrightarrow \xi'^-$ and then make variable changes
$\xi'^- \rightarrow \xi'^- -y^-$ and $\eta'^- \rightarrow \eta'^- -y^-$. We also assume the translational invariance of the matrix element $\langle N|F_{\perp}^{\rho+}(\xi'^- +y^-)
F_{\perp\rho}^{\ +}(\eta'^- + y^-)|N\rangle =\langle N|F_{\perp}^{\rho+}(\xi'^-)
F_{\perp\rho}^{\ +}(\eta'^-)|N\rangle$ and $\rho_N^A({\xi'}_N^- +y^-)\approx \rho_N^A({\xi'}_N^-)$. The final result in Eq.~\eqref{EQ17} is obtained with another variable change $z^- \rightarrow z^- -y^-$.

Note that there is a sign difference between contributions from the left and right-cut diagrams.  The net sum of these two contributions gives,
\begin{widetext}
\begin{eqnarray}
\langle \vec k_\perp^4\varphi_{\perp}^{(2)A}\rangle_1(x)&=&\frac{1}{2}\left[
\langle \vec k_\perp^4\varphi_{\perp}^{(2,L)A}\rangle_1(x) +\langle \vec k_\perp^4\varphi_{\perp}^{(2,R)A}\rangle_1(x) \right] \notag\\
%&=&A\int\frac{dy^-}{2\pi}e^{ixp^+y^-}\langle N|\bar{\psi}(0)
%\frac{\gamma^+}{2}\psi(y^-)|N\rangle\notag\\
%&&\times\int_0^{\infty}d\xi^-\int_{\xi^-}^{\infty}d\eta^-
%g^2\langle N|F_{\perp}^{\sigma+}(\xi^-)
%F_{\perp\sigma}^{\ +}(\eta^-)|N\rangle\notag\\
%&&\times\int p^+dz^-i\theta(z^-)\left[\int_0^{z^-}d\xi'^-
%\int_0^{\xi'^-}d\eta'^--\int_0^{z^-}d\xi'^-
%\int_{\xi'^-}^{z^-}d\eta'^-\right]\notag\\
%&&\times2g^2\langle N|F_{\perp}^{\rho+}(\xi'^-)
%F_{\perp\rho}^{\ +}(\eta'^-)|N\rangle\notag\\
&&\hspace{-0.5in}\simeq Af_q^N(x) \Delta_{2F} 
 \int p^+dz^- i\theta(z^-)\int_0^{z^-}\hspace{-8pt}d\xi'^- \hspace{-4pt} \int_0^{\xi'^-}\hspace{-8pt}d\eta'^- 
  g^2\langle N|\left[F_{\perp}^{\rho+}(\xi'^-),
F_{\perp\rho}^{\ +}(\eta'^-)\right]|N\rangle \frac{\rho_N^A({\xi'}_N^-)}{2p^-}  =0,
\label{EQ18}
\end{eqnarray}
\end{widetext}
that contains a commutator of the gluon field strength tensor which should vanish on the light-cone \cite{Luo:1992fz, Luo:1994np}.
Therefore, $\langle \vec k_\perp^4\varphi_{\perp}^{(2)A}\rangle(x)$ has no leading contribution with $A^{5/3}$ nuclear 
enhancement due to the cancellation between left and right-cut diagrams. 

Using similar techniques one finds that the contributions from the third term of Eq.~\eqref{EQ15} also vanishes due to
cancellation between left and right-cut diagrams. The only remaining contribution with nuclear enhancement is the second 
term in Eq.~\eqref{EQ15} which can be cast in the form,
\begin{equation}
\langle \vec k_\perp^4\varphi_{\perp}^{(2)A}\rangle(x)=A\psi_\perp^{(2)N}(x)\Delta_{2F}  +{\cal O}(A),
\end{equation}
where
\begin{widetext}
\begin{eqnarray}
\psi_{\perp}^{(2,L)N}(x)&=&\int \frac{dy^-}{2\pi}p^+dz^-e^{ixp^+y^-}i\theta(z^-)
\langle N|\bar{\psi}(0)\frac{\gamma^+}{2}D_\perp^2(z^-)\psi(y^-)|N\rangle
\label{EQ19}\\
\psi_{\perp}^{(2,R)N}(x)&=&\int \frac{dy^-}{2\pi}p^+dz^-e^{ixp^+y^-} (-i)\theta(z^- - y^-)
\langle N|\bar{\psi}(0)\frac{\gamma^+}{2}D_\perp^2(z^-)\psi(y^-)|N\rangle, \\
\label{EQ20}
\psi_\perp^{(2)N}(x)&=&\frac{1}{2}\left[ \psi_\perp^{(2,L)N}(x)+\psi_\perp^{(2,R)N}(x)\right]
=ip^+\int \frac{dy^-}{4\pi}\int_0^{y^-}\hspace{-8pt}dz^-e^{ixp^+y^-}
\langle N|\bar{\psi}(0)\frac{\gamma^+}{2}D_\perp^2(z^-)\psi(y^-)|N\rangle.
\label{EQ202}
\end{eqnarray}
\end{widetext}
are twist-four parton correlation functions for a nucleon state. We have omitted the gauge links in the above expressions for brevity.
%Note that the integration over $z^-$ in Eq.~(\ref{EQ202}) is limited to $y^-$ which is in turn confined to the size of a single nucleon.
%The twist-four parton correlation function $\psi_\perp^{(2)A}(x) \propto A \psi_\perp^{(2)N}(x)$ does not have extra nuclear enhancement.
One can similarly find that
\begin{equation}
\langle \vec k_\perp^4\tilde\varphi_{\perp}^{(2)A}\rangle(x) = A \tilde\psi_\perp^{(2)N}(x)\Delta_{2F} +{\cal O}(A).
\end{equation}
The definition of $\tilde{\psi}_{\perp}^{(2)N}(x)$ is similar to $\tilde{\psi}_{\perp}^{(2)N}(x)$. Both are given in the appendix.

%%%%%%%%%%%%%%%%%%%%%%%%%%%%%%%%%%%%%%%%%%%%%%%%%%%%%%%%%%%
\section{\label{sec4} Transverse Momentum Broadening}

In the previous section, the $\vec k_\perp^2$-weighted cross section should also contain contributions that are not enhanced by the nuclear size, including those that contain parton correlations inside a single nucleon. These contributions are essentially the same cross section for a single nucleon target multiplied by the atomic number $A$. We should subtract these contributions to obtain the final result for the nuclear enhanced $k_\perp^2$-weighted differential cross section,
\begin{eqnarray}
 &&\hspace{-0.4in}\frac{d \langle \vec k_\perp^2 \sigma^A\rangle}{dx_Bdy} -  A\frac{d \langle \vec k_\perp^2 \sigma^N\rangle}{dx_Bdy}  \notag\\
&&\hspace{-0.1in}\simeq \frac{2\pi\alpha_{\rm em}^2e_q^2}{Q^2y} A \Delta_{2F} \bigg\{[1+(1-y)^2]f_q^N(x_B) \notag\\
&&\hspace{-0.1in}-\frac{2[1+(1-y)^2]x_B}{Q^2}\left[
\psi_{\perp}^{(2)N}(x_B)-\tilde{\psi}_{\perp}^{(2)N}(x_B) \right] \notag\\
&&\hspace{-0.1in}+\frac{16(1-y)x_B^2M^2}{Q^2}f_{q(-)}^N(x_B)\bigg\}.
\label{EQ33}
\end{eqnarray}

The averaged transverse momentum broadening is defined as,
\begin{equation}
\triangle \langle p_\perp^2\rangle\equiv
\frac{d \langle \vec k_\perp^2 \sigma^A\rangle}{d\sigma^A} -  \frac{d \langle \vec k_\perp^2 \sigma^N\rangle}{d\sigma^N},
\end{equation}
where the transverse momentum integrated cross section can be obtained from Eq.~\eqref{EQ01},
\begin{eqnarray}
\frac{d\sigma^A}{dx_Bdy}&=&\frac{2\pi\alpha_{\rm em}^2e_q^2}{Q^2y}
\bigg\{[1+(1-y)^2]f_q^A(x_B) \notag\\
&&\hspace{-0.4in}-2[1+(1-y)^2]\frac{x_B}{Q^2}[\psi_{\perp}^{(2)A}(x_B)
-\tilde{\psi}_{\perp}^{(2)A}(x_B)] \notag\\
&&\hspace{-0.4in}+16(1-y)\frac{x_B^2M^2}{Q^2}f_{q(-)}^A(x_B)\bigg\} .
\label{EQ39}
\end{eqnarray}
The above expression for transverse-momentum-integrated cross section is suitable for both nuclear and nucleon targets.
The definition of twist-four parton correlation functions $\psi_{\perp}^{(2)A}(x_B)$ and $\tilde{\psi}_{\perp}^{(2)A}(x_B)$ (see the appendix)
are also suitable for both nuclei and nucleons.  Because of the cancellation between contributions from right and left-cut diagrams, the
interaction with the soft gluon field is limited within the size of a single nucleon [the integration over $z^-$ in Eq.~(\ref{EQ202})
for $\psi_{\perp}^{(2)A}(x_B)$ is limited to the size $y^-$ of the nucleon, for example]. Therefore, $\psi_{\perp}^{(2)A}(x_B)$ and $\tilde{\psi}_{\perp}^{(2)A}(x_B)$  have no nuclear enhancement, similarly as $f_q^A(x_B)$ and $f_{q(-)}^A(x_B)$.

Including only the leading nuclear enhancement in the $\vec k_\perp^2$-weighted and $\vec k_\perp$-integrated differential cross sections and expanding both in terms of power-corrections, one has
\begin{widetext}
\begin{eqnarray}
\triangle\langle p_\perp^2\rangle
\simeq \Delta_{2F} \frac{Af_q^N(x_B)}{f_q^A(x_B)} \bigg\{
1&+&\frac{16(1-y)}{[1+(1-y)^2]}\frac{x_B^2M^2}{Q^2}
\left[\frac{f_{q(-)}^N(x_B)}{f_q^N(x_B)}
-\frac{f_{q(-)}^A(x_B)}{f_q^A(x_B)}\right]\notag\\
&-&\frac{2x_B}{Q^2}\left[\frac{\psi_{\perp}^{(2)N}(x_B)}
{f_q^N(x_B)}-\frac{\tilde{\psi}_{\perp}^{(2)N}(x_B)}{f_q^N(x_B)}
-\frac{\psi_{\perp}^{(2)A}(x_B)}{f_q^A(x_B)}
+\frac{\tilde{\psi}_{\perp}^{(2)A}(x_B)}{f_q^A(x_B)} \right]\bigg\}.
\end{eqnarray}
\end{widetext}
Since all the collinear nuclear parton distribution and correlation functions, $f_q^A(x_B)$, $f_{q(-)}^A(x_B)$, $\psi_\perp^{(2)A}(x_B)$ and
$\tilde\psi_\perp^{(2)A}(x_B)$, have no extra nuclear enhancement beside their linear dependence on $A$, their ratios $f_{q(-)}^A(x_B)/f_q^A(x_B)$, 
$\psi_\perp^{(2)A}(x_B)/f_q^A(x_B)$ and $\tilde\psi_\perp^{(2)A}(x_B)/f_q^A(x_B)$, also do not have any nuclear enhancement. Therefore, the power corrections in the above transverse momentum broadening are not enhanced by the nuclear size. The finite power corrections to the transverse momentum broadening depend, however, on the difference between the collinear parton distribution and correlation functions in a nucleus and that in a free nucleon. If the nuclear parton distribution and correlation functions are just the sums of those of free nucleons, {\it i.e.}, $f_q^A(x_B)=Af_q^N(x_B)$, $f_{q(-)}^A(x_B)=Af_{q(-)}^N(x_B)$, $\psi_\perp^{(2)A}(x_B)=A\psi_\perp^{(2)N}(x_B)$,
$\tilde\psi_\perp^{(2)A}(x_B)=A\tilde\psi_\perp^{(2)N}(x_B)$, the above power corrections will all vanish. Therefore, the power corrections to the transverse momentum broadening are sensitive to the nuclear modifications of the collinear parton distribution and correlation functions.

%%%%%%%%%%%%%%%%%%%%%%%%%%%%%%%%%%%%%%%%%%%%%%%%%%%%%%%%%%%%%%%%
\section{Conclusion}
\label{sec5}

In conclusion, we use the differential TMD cross section of SIDIS up to twist-four contributions to calculate the transverse momentum
broadening of the struck quark up to the first power corrections in $1/Q^2$.  The $\vec k_\perp^2$-weighted differential cross section depends on higher-twist collinear matrix elements of nuclear states up to twist-six which are shown to factorize approximately into products of jet transport parameter and collinear parton distribution and correlation functions. Because of the cancellation between left and right-cut diagrams, these higher-twist matrix elements are shown to have no extra nuclear enhancement in addition to that involved with the jet transport parameter. The final transverse momentum broadening of the struck quark is shown to be proportional to the leading twist result given by a path-integral of the jet transport parameter. The next power corrections are found to have no additional nuclear enhancement. The coefficients of these power corrections are determined by the nuclear modification of the collinear parton distribution and correlation functions.

Such power corrections to the transverse momentum broadening might be important for extracting the jet transport parameter from experimental data.  Understanding of these power corrections is also important to the study of the scale evolution of the jet transport parameter according to the QCD evolution equation from the NLO higher-twist contributions to the SIDIS cross section \cite{Kang:2013raa}, especially at intermediate values of $Q^2$.

\section*{Acknowledgments}
We thank Dr.~Hongxi Xing and Dr.~Feng Yuan for helpful discussions
and useful comments. This work is supported in part by Natural Science Foundation of China under grant Nos. 11221504, 11105137 and 11475104, Chinese Ministry of Science and Technology under Grant No. 2014DFG02050, the Major State Basic Research Development Program in
China (Grant Nos. 2014CB845406 and 2014CB845404), by the Director, Office of Energy Research, Office of High Energy and Nuclear Physics, Division of Nuclear Physics, of the U.S. Department of Energy under Contract No. DE- AC02-05CH11231 and within the framework of the JET Collaboration. J.H.G is also supported by CCNU-QLPL Innovation Fund (QLPL2014P01).
%%%%%%%%%%%%%%%%%%%%%%%%%%%%%%%%%%%%%%%%%%%%%%%%%%%%%%%%%
\appendix

\section{Definitions of the matrix elements}

%\section*{Appendix: Definitions of the matrix elements}
The TMD parton distribution and correlation functions in terms of matrix elements of parton field operators that we use in the differential cross
sections of SIDIS in Eq.~(\ref{EQ01}) are defined as follows  \cite{Song:2010pf},
\begin{eqnarray}
f_q^A(x,\vec k_\perp)&=&\int\frac{dy^-d^2y_{\perp}}{(2\pi)^3}
e^{ixp^+y^- - i\vec{k}_{\perp}\cdot\vec{y}_{\perp}} \notag\\
&\times& \langle A|\bar{\psi}(0)\frac{\gamma^+}{2}\mathcal{L}(0;y)
\psi(y)|A\rangle,\\
\label{EQA01}
%\end{eqnarray}
%
%\begin{eqnarray}
f^A_{q\perp}(x,\vec k_\perp)&=&\int\frac{p^+dy^-d^2y_{\perp}}{(2\pi)^3}
e^{ixp^+y^- -i\vec{k}_{\perp}\cdot\vec{y}_{\perp}}  \notag\\
&\times& \langle A|\bar{\psi}(0)\frac{\slashed{k}_{\perp}}{2\vec k_\perp^2}
\mathcal{L}(0;y)\psi(y)|A\rangle,\\
\label{EQA02}
%\end{eqnarray}
%
%\begin{eqnarray}
f^A_{q(-)}(x,\vec k_\perp)&=&\frac{1}{M^2}\int\frac{p^+dy^-d^2y_{\perp}}
{(2\pi)^3}e^{ixp^+y^- - i\vec{k}_{\perp}\cdot\vec{y}_{\perp}}  \notag\\
&\times& \langle A|\bar{\psi}(0)\frac{\slashed{p}}{2}\mathcal{L}(0;y)
\psi(y)|A\rangle.
\label{EQA03}
\end{eqnarray}
The corresponding collinear ($\vec k_\perp$-integrated) parton distribution and correlation functions, $f_q^A(x)$ and $f_{q(-)}^A(x)$, can be obtained from the above by integration over $\vec k_\perp$. 

The following high-twist TMD parton distribution functions in terms of the matrix elements from both left and right cut diagrams
in Fig.~\ref{fig01}:

\begin{eqnarray}
\varphi_{\perp2}^{(i)A} & \equiv& \frac{1}{2}\left( \varphi_{\perp2}^{(i,L)N}+\varphi_{\perp2}^{(i,R)A}\right), (i=1,2),\\
\tilde{\varphi}_{\perp2}^{(i)A} &\equiv& \frac{1}{2}\left( \tilde{\varphi}_{\perp2}^{(i,L)A}+\tilde{\varphi}_{\perp2}^{(i,R)A} \right), (i=1,2),
\label{EQA04}
\end{eqnarray}
where,
\begin{widetext}
\begin{eqnarray}
\varphi_{\perp2}^{(1,L)A}&=&\int\frac{p^+dy^-d^2y_{\perp}}{(2\pi)^3}
e^{ixp^+y^--i\vec{y}_{\perp}\cdot\vec{k}_{\perp}}
\frac{(2k_{\perp\alpha}k_{\perp\rho}-k_\perp^2d_{\rho\alpha})}
{\vec k_\perp^4}\langle A|\bar{\psi}(0)\frac{\gamma_{\alpha}}{2}
D_{\rho}(0)\mathcal{L}(0;y)\psi(y)|A\rangle,
\label{EQA06}\\
\tilde{\varphi}_{\perp2}^{(1,L)A}&=&\int\frac{p^+dy^-d^2y_{\perp}}
{(2\pi)^3}e^{ixp^+y^- -i\vec{y}_{\perp}\cdot\vec{k}_{\perp}}
\frac{(-ik_{\perp\{\alpha}\epsilon_{\perp\rho\}\gamma}
k_{\perp}^{\gamma})}{\vec k_\perp^4}\langle A|\bar{\psi}(0)
\frac{\gamma_5\gamma_{\alpha}}{2}D_{\rho}(0)\mathcal{L}(0;y)
\psi(y)|A\rangle,
\label{EQA07}\\
\varphi_{\perp2}^{(1,R)A}&=&\int\frac{p^+dy^-d^2y_{\perp}}
{(2\pi)^3}e^{ixp^+y^--i\vec{y}_{\perp}\cdot\vec{k}_{\perp}}
\frac{(2k_{\perp\alpha}k_{\perp\rho}-k_\perp^2d_{\rho\alpha})}
{\vec k_\perp^4}\langle A|\bar{\psi}(0)\frac{\gamma_{\alpha}}{2}
\mathcal{L}(0;y)D_{\rho}(y)\psi(y)|A\rangle,
\label{EQA08}\\
\tilde{\varphi}_{\perp2}^{(1,R)A}&=&
\int\frac{p^+dy^-d^2y_{\perp}}{(2\pi)^3}
e^{ixp^+y^--i\vec{y}_{\perp}\cdot\vec{k}_{\perp}}
\frac{(-ik_{\perp\{\alpha}\epsilon_{\perp\rho\}\gamma}
k_{\perp}^{\gamma})}{\vec k_\perp^4}\langle A|\bar{\psi}(0)
\frac{\gamma_5\gamma_{\alpha}}{2}\mathcal{L}(0;y)D_{\rho}(y)
\psi(y)|A\rangle,
\label{EQA09}
\end{eqnarray}
\begin{eqnarray}
\varphi_{\perp}^{(2,L)A}&=&\int\frac{dx_2}{x_2-x-i\epsilon}
\frac{dy^-d^2y_{\perp}}{(2\pi)^3}\frac{p^+dz^-}{2\pi}
e^{ix_2p^+z^-+ixp^+(y^--z^-)-i\vec{k}_{\perp}\cdot\vec{y}_{\perp}}
\frac{-d^{\rho\sigma}}{k_\perp^2}\notag\\
&&\times\langle A|\bar{\psi}(0)\frac{\gamma^+}{2}
\mathcal{L}(0;z^-,0_{\perp})D_{\rho}(z^-,0_{\perp})
D_{\sigma}(z^-,0_{\perp})\mathcal{L}(z^-,0_{\perp};y)
\psi(y)|A\rangle,\label{EQA10}\\
\tilde{\varphi}_{\perp}^{(2,L)A}&=&\int\frac{dx_2}{x_2-x-i\epsilon}
\frac{dy^-d^2y_{\perp}}{(2\pi)^3}\frac{p^+dz^-}{2\pi}
e^{ix_2p^+z^-+ixp^+(y^--z^-)-i\vec{k}_{\perp}\cdot\vec{y}_{\perp}}
\frac{-i\epsilon_{\perp}^{\rho\sigma}}{k_\perp^2}\notag\\
&&\times\langle A|\bar{\psi}(0)\frac{\gamma_5\gamma^+}{2}
\mathcal{L}(0;z^-,0_{\perp})D_{\rho}(z^-,0_{\perp})
D_{\sigma}(z^-,0_{\perp})\mathcal{L}(z^-,0_{\perp};y)
\psi(y)|A\rangle,\label{EQA11}\\
\varphi_{\perp}^{(2,R)A}&=&\int\frac{dx_1}{x_1-x+i\epsilon}
\frac{dy^-d^2y_{\perp}}{(2\pi)^3}\frac{p^+dz^-}{2\pi}
e^{ixp^+z^-+ix_1p^+(y^--z^-)-i\vec{k}_{\perp}\cdot\vec{y}_{\perp}}
\frac{-d^{\rho\sigma}}{k_{\perp}^2}\notag\\
&&\times\langle A|\bar{\psi}(0)\frac{\gamma^+}{2}
\mathcal{L}(0;z^-,y_{\perp})D_{\rho}(z^-,y_{\perp})
D_{\sigma}(z^-,y_{\perp})\mathcal{L}(z^-,y_{\perp};y)
\psi(y)|A\rangle,\label{EQA12}\\
\tilde{\varphi}_{\perp}^{(2,R)A}&=&\int\frac{dx_1}{x_1-x+i\epsilon}
\frac{dy^-d^2y_{\perp}}{(2\pi)^3}\frac{p^+dz^-}{2\pi}
e^{ixp^+z^-+ix_1p^+(y^--z^-)-i\vec{k}_{\perp}\cdot\vec{y}_{\perp}}
\frac{-i\epsilon_{\perp}^{\rho\sigma}}{k_{\perp}^2}\notag\\
&&\times\langle A|\bar{\psi}(0)\frac{\gamma_5\gamma^+}{2}
\mathcal{L}(0;z^-,y_{\perp})D_{\rho}(z^-,y_{\perp})
D_{\sigma}(z^-,y_{\perp})\mathcal{L}(z^-,y_{\perp};y)\psi(y)|A\rangle,
\label{EQA13}
\end{eqnarray}
\end{widetext}
where $d^{\mu\nu}=g^{\mu\nu}-\bar{n}^{\mu}n^{\nu}-\bar{n}^{\nu}n^{\mu}$, $\epsilon_{\perp}^{\mu\nu}=\epsilon^{\mu\nu\rho\sigma}\bar n_{\rho} n_\sigma$ and $\epsilon^{\mu\nu\rho\sigma}$ the anti-symmetric unit tensor. The four-vector of the transverse momentum is defined $k_\perp=(0,0,\vec k_\perp)$.

In the calculation of the $\vec k_\perp^2$-weighted differential cross section, the following collinear parton distribution and correlation functions are used.
\begin{widetext}
\begin{eqnarray}
\langle \vec k_\perp^2 f_q^A\rangle (x)&=&\int d^2\vec k_\perp \vec{k}_{\perp}^2 f_q^A(x,\vec k_\perp)\notag\\
&=&\int\frac{dy^-d^2y_{\perp}d^2\vec k_\perp}{(2\pi)^3}
e^{ixp^+y^--i\vec{k}_{\perp}\cdot\vec{y}_{\perp}} \vec{k}_{\perp}^2 \langle A|\bar{\psi}(0)\frac{\gamma^+}{2}
\mathcal{L}(0;y)\psi(y)|A\rangle,
\label{EQA14}
\end{eqnarray}
\begin{eqnarray}
\langle \vec k_\perp^2 f^A_{q(-)}\rangle(x)&=&\int d^2\vec k_\perp\vec{k}_{\perp}^2  f^A_{q(-)}(x,\vec k_\perp)\notag\\
&=&\frac{p^+}{M^2}\int\frac{dy^-d^2y_{\perp}d^2\vec k_\perp}{(2\pi)^3}
e^{ixp^+y^- - i\vec{k}_{\perp}\cdot\vec{y}_{\perp}}
 \vec{k}_{\perp}^2\langle A|\bar{\psi}(0)\frac{\slashed{p}}{2}
\mathcal{L}(0;y)\psi(y)|A\rangle,
\label{EQA15}
\end{eqnarray}
%
%\Psi_{\perp}^{(2)N}(x)&\equiv&\frac{1}{2}\left[
%\Psi_{\perp}^{(2,L)N}(x)+\Psi_{\perp}^{(2,R)N}(x)\right],
%\qquad \tilde{\Psi}_{\perp}^{(2)N}(x)
%\equiv\frac{1}{2}\left[\tilde{\Psi}_{\perp}^{(2,L)N}(x)
%+\tilde{\Psi}_{\perp}^{(2,R)N}(x)\right]
%\label{EQA16}
%\end{eqnarray}
%The definitions for $\Psi_{\perp}^{(2,L)N}$, $\Psi_{\perp}^{(2,R)N}$,
%$\tilde{\Psi}_{\perp}^{(2,L)N}$ and $\tilde{\Psi}_{\perp}^{(2,R)N}$
%in Eq.~\eqref{EQA16} are
%\begin{eqnarray}

\begin{eqnarray}
\langle \vec k_\perp^4 \varphi_{\perp}^{(2,L)A}\rangle (x)&=&
\int\frac{d^2\vec k_\perp dx_2}{x_2-x-i\epsilon}
\frac{dy^-d^2y_{\perp}}{(2\pi)^3}\frac{p^+dz^-}{2\pi}
e^{ix_2p^+z^-+ixp^+(y^--z^-)-i\vec{k}_{\perp}\cdot\vec{y}_{\perp}}
\left(d^{\rho\sigma}\vec k_\perp^2\right)\notag\\
&&\times\langle A|\bar{\psi}(0)\frac{\gamma^+}{2}
\mathcal{L}(0;z^-,0_{\perp})D_{\rho}(z^-,0_{\perp})
D_{\sigma}(z^-,0_{\perp})\mathcal{L}(z^-,0_{\perp};y)
\psi(y)|A\rangle,
\label{EQA17}
\end{eqnarray}
\begin{eqnarray}
\langle \vec k_\perp^4 \varphi_{\perp}^{(2,R)A}\rangle(x)&=&
\int\frac{d^2\vec k_\perp dx_1}{x_1-x+i\epsilon}
\frac{dy^-d^2y_{\perp}}{(2\pi)^3}\frac{p^+dz^-}{2\pi}
e^{ixp^+z^-+ix_1p^+(y^--z^-)-i\vec{k}_{\perp}\cdot\vec{y}_{\perp}}
\left(d^{\rho\sigma}\vec k_\perp^2\right)\notag\\
&&\times\langle A|\bar{\psi}(0)\frac{\gamma^+}{2}
\mathcal{L}(0;z^-,y_{\perp})D_{\rho}(z^-,y_{\perp})
D_{\sigma}(z^-,y_{\perp})\mathcal{L}(z^-,y_{\perp};y)\psi(y)
|A\rangle,
\label{EQA18}
\end{eqnarray}
\begin{eqnarray}
\langle \vec k_\perp^4 \tilde{\varphi}_{\perp}^{(2,L)A}\rangle(x)&=&
\int\frac{d^2\vec k_\perp dx_2}{x_2-x-i\epsilon}
\frac{dy^-d^2y_{\perp}}{(2\pi)^3}\frac{p^+dz^-}{2\pi}
e^{ix_2p^+z^-+ixp^+(y^--z^-)-i\vec{k}_{\perp}\cdot\vec{y}_{\perp}}
\left(i\epsilon_{\perp}^{\rho\sigma}\vec k^2_\perp\right)
\notag\\
&&\times\langle A|\bar{\psi}(0)\frac{\gamma_5\gamma^+}{2}
\mathcal{L}(0;z^-,0_{\perp})D_{\rho}(z^-,0_{\perp})
D_{\sigma}(z^-,0_{\perp})\mathcal{L}(z^-,0_{\perp};y)
\psi(y)|A\rangle,
\label{EQA19}
\end{eqnarray}
\begin{eqnarray}
\langle \vec k_\perp^4 \tilde{\varphi}_{\perp}^{(2,R)A}\rangle(x)&=&
\int\frac{d^2\vec k_\perp dx_1}{x_1-x+i\epsilon}
\frac{dy^-d^2y_{\perp}}{(2\pi)^3}\frac{p^+dz^-}{2\pi}
e^{ixp^+z^-+ix_1p^+(y^--z^-)-i\vec{k}_{\perp}\cdot\vec{y}_{\perp}}
\left(i\epsilon_{\perp}^{\rho\sigma}\vec k^2_\perp\right)
\notag\\
&&\times\langle A|\bar{\psi}(0)\frac{\gamma_5\gamma^+}{2}
\mathcal{L}(0;z^-,y_{\perp})D_{\rho}(z^-,y_{\perp})
D_{\sigma}(z^-,y_{\perp})\mathcal{L}(z^-,y_{\perp};y)\psi(y)
|A\rangle.
\label{EQA20}
\end{eqnarray}

In the calculation of the $\vec k_\perp$-integrated differential cross section, one also encounters the following collinear 
parton correlation functions,
\begin{eqnarray}
\psi_{\perp}^{(2)A}(x)&=& \frac{1}{2}\left[\langle \vec k_\perp^2 \varphi_{\perp}^{(2,L)A}\rangle(x) 
+ \langle \vec k_\perp^2 \varphi_{\perp}^{(2,R)A}\rangle(x)\right] \notag\\
&=&ip^+\int \frac{dy^-}{4\pi}\int_0^{y^-} dz^-e^{ixp^+y^-} \langle A|\bar{\psi}(0)\frac{\gamma^+}{2}
\mathcal{L}(0;z^-)D_\perp^2(z^-) \mathcal{L}(z^-;y^-)\psi(y^-)|A\rangle,\\
 \label{EQA25}
\tilde{\psi}_{\perp}^{(2)A}(x)&=& \frac{1}{2}\left[\langle \vec k_\perp^2 \tilde{\varphi}_{\perp}^{(2,L)A}\rangle(x)
+\langle \vec k_\perp^2 \tilde{\varphi}_{\perp}^{(2,R)A}\rangle(x)\right] \notag\\
&=&-p^+\int \frac{dy^-}{4\pi}\int_0^{y^-} dz^-e^{ixp^+y^-} 
\epsilon_{\perp}^{\rho\sigma}\langle A|\bar{\psi}(0)\frac{\gamma^5\gamma^+}{2}
\mathcal{L}(0;z^-)D_{\rho}(z^-)D_{\sigma}(z^-)
\mathcal{L}(z^-;y^-)\psi(y^-)|A\rangle .
\label{EQA28}
\end{eqnarray}
\end{widetext}
Since nucleons are color singlet states, the quark field operators in the above matrix elements must operate on a single nucleon
inside the nucleus. This limits the range of the coordinate $y^-$ to the size of a nucleon, which also limits the range of $z^-$ integration. Therefore, the above matrix elements should be proportional to the atomic number of the nucleus and do not have any extra nuclear enhancement.

All the matrix elements defined above for nuclear states are also valid for a nucleon.  The $\vec k_\perp^4$-weight-integrated nuclear parton correlation functions $\langle \vec k_\perp^4 \varphi_{\perp}^{(2)A}\rangle(x)$ and $\langle \vec k_\perp^4 \tilde{\varphi}_{\perp}^{(2)A}\rangle(x)$ are also related to the parton correlation functions in Eqs.~(A20) and (\ref{EQA28}) for a  nucleon state as shown in Sec. \ref{sec3}.

%\bibliographystyle{h-physrev5}
%\bibliography{ref}

\end{document}